\documentclass[twocolumn,hidelinks]{article}

\usepackage{amsmath} 
\usepackage{amssymb} 

\newcommand{\eq}[1]{{eq.~(\ref{#1})}}
\newcommand{\er}[1]{{(\ref{#1})}}
\newcommand{\Tr}[1]{\mathrm{Tr}(#1)}
\newcommand{\trfu}[1]{U #1 U^{-1}}
\begin{document}
                   
\title{\bf \Large Finding room for antilinear terms in the Hamiltonian}

\author{Michael Eisele \\ \em \small
Department of Mathematics and Physics M5, 
\\ \em \small Technical University of Munich, D-85748 Garching, Germany 
\\ \small email: mbjeisele@gmail.com 
}

\twocolumn[ 
  \begin{@twocolumnfalse}
    \maketitle
    \begin{abstract} \normalsize
    Although the Hamiltonian in quantum physics has to be a linear operator, it is possible to make quantum systems behave as if their Hamiltonians contained antilinear (i.e., semilinear or conjugate-linear) terms. For any given quantum system, another system can be constructed that is physically equivalent to the original one. It can be designed, despite the Wightman reconstruction theorem, so that antilinear operators in the original system become linear operators in the new system. Under certain conditions, these operators can then be added to the new Hamiltonian. The new quantum system has some unconventional features, a hidden degeneracy of the vacuum and a subtle distinction between the Hamiltonian and the observable of energy, but the physical equivalence guarantees that its states evolve like those in the original system and that corresponding measurements produce the same results. The same construction can be used to make time-reversal linear.
\\~
\end{abstract}

\end{@twocolumnfalse}
]

\section{Introduction} \label{s1}
Fundamental principles of quantum physics require the Hamiltonian to be a linear operator. Wigner's unitary-antiunitary theorem \cite{wigner1931}, in particular, tells us that symmetries can be described, up to a phase, by unitary or antiunitary operators. This implies that generators of continuous symmetries should be linear operators (e.g., \cite[chapter 2]{weinberg1995quantum}). As the Hamiltonian is the generator of temporal displacements, it should also be linear. 

For certain applications, which we will discuss in section~\ref{s5}, it would, however, help if some terms in the Hamiltonian could be antilinear. Such antilinear operators, which are also called conjugate-linear or semilinear, are known from time-reversal, commute with real numbers, but anticommute with the imaginary unit $i$ (e.g., \cite{sharma1988semilinear}). Here we show a general procedure to make quantum systems behave as if their Hamiltonian contained antilinear terms. For any given quantum system, we can construct another physically equivalent system where such antilinear terms become linear and where they can be added to the Hamiltonian. 

This result may surprise, because the Wightman reconstruction theorem tells us that any quantum field theory can be reconstructed, up to a unitary transformation, from its correlation functions \cite{streater1964pct}, and because a unitary transformation cannot turn antilinear terms into linear ones. However, the Wightman theorem applies only to systems where the vacuum state is unique (up to a complex phase). We will design the new quantum system so that its vacuum states are degenerate (encompassing at least two orthogonal vectors of the Hilbert space). This degeneracy is hidden, because of the physical equivalence to the original quantum system, but it provides room for the new system to differ from the original one by more than just a unitary transformation.

To construct the new quantum system, given the original one, we proceed in two steps. Given any quantum system, called system A, we first construct a system B, with degeneracy, and then a system C where antilinear operators turn linear. The first step, in section~\ref{s2}, is almost trivial. By taking the direct sum of the Hilbert space with itself, and by lifting the observables and the Hamiltonian to the resulting space, we introduce a twofold degeneracy of the vacuum and of other states. This new system B has more possible states than the original one, but its observables fulfill a certain constraint which makes some states indistinguishable. To prove physical equivalence, we will identify corresponding states in both systems, show that they evolve in parallel, and show that they produce the same results during quantum measurements. 

The second step, in section~\ref{s3}, is less obvious. In system B we introduce an operator $j$ that toggles between degenerate states. It somewhat resembles the imaginary unit $i$ because:
\begin{equation} \label{eIN3}
j^2 = -1 \, ; \;\; j^\dag = -j
\end{equation}
(see \eq{eVV3c}). It was motivated by a similar operator $\breve{J}$ known from quantum physics on real Hilbert spaces \cite{stueckelberg1959field,stueckelberg1960quantum}, but, unlike $\breve{J}$, it exists alongside $i$. We then construct system C by replacing some occurrences of $i$ with $j$. The Schr\"odinger and the von Neumann equation retain their usual form, with $i$ instead of $j$, and vectors differing by a complex phase $\exp(i a)$ will still belong to the same physical state, but in observables and in the Hamiltonian $j$ may replace $i$. Physical equivalence can again be shown, so that the degeneracy remains hidden, but antilinear operators, which anticommute with $i$, may turn into linear operators that anticommute with $j$.

Section~\ref{s5} shows how this construction makes room for antilinear terms in the Hamiltonian. Antilinear operators $H^A_2$, which cannot be directly added to the original Hamiltonian $H^A$, become linear operators $H^C_2$, in system C, and may be added to the new Hamiltonian $H^C$. States then behave as if they would evolve under $H^A + H^A_2$, which is not linear, even though they actually evolve under the linear Hamiltonian $H^C + H^C_2$.

Part of this construction is known, especially for the Dirac equation, from the field of quantum simulations. It has been shown before that such a system C can simulate any state of system A and that antilinear operators in system A become linear operators in system C \cite{casanova2011quantum}. Here it is shown that the converse also holds, with system A being able to simulate any state of system C, so that both systems can be regarded as physically equivalent. That is, each state in one system corresponds to (at least) one state in the other system, corresponding states evolve in parallel, and they produce the same results, with the same probabilities, when corresponding observables are measured. While mathematically different, such systems cannot be told apart by physical observations.

We will also take a closer look at mixed states, compare the measurement process in both systems, and address a feature of system C which is unconventional: By construction, the observable of energy of system C will differ from its Hamiltonian (section~\ref{s33b}). Both operators will, however, obey the usual rules, such as being self-adjoint, and their difference will be rather subtle. Due to the physical equivalence, of systems A and C, this feature seems permissible. 

As linear Hamiltonians have been very successful at describing physics, the importance of adding antilinear terms is not self-evident, so section~\ref{s5} also takes a few steps towards possible applications. It indicates how these results may be used to turn an antilinear time-reversal operator $T^A$ into a linear time-reversal operator $T^C$, to replace right-handed Weyl spinors by left-handed ones, or to make room for continuous symmetries with antilinear generators. The appendix summarizes useful properties of sums of linear and antilinear operators and their adjoints. For simplicity, we will not attempt to prove all mathematical results in full generality. Some relations are proven only for finite Hilbert spaces, even though they may hold on infinite Hilbert spaces as well.

\section{From system A to B} \label{s2}
Let us consider any quantum system. We call it system A, its Hamiltonian $H^A$, and its Hilbert space $\mathcal{H}^A$. The Hamiltonian $H^A$ is linear, and, for simplicity, we assume that it does not explicitly depend on time. The vacuum $\Theta^A$, the state of lowest energy, is assumed to be unique (although the same procedure would work if the vacuum was already degenerate). We also assume that any positive semidefinite, self-adjoint operator $\rho$ of trace 1 constitutes a possible density operator of this system (although the results may be generalized to systems with superselection rules). We do not assume that every self-adjoint operator constitutes an observable, but let system A be characterized by a given set of observables $O^A$.

The goal of this section is to construct another system B, with Hamiltonian $H^B$ and Hilbert space $\mathcal{H}^B$, which is physically equivalent to the original one but has a hidden, twofold degeneracy of the vacuum. This will demonstrate how two equivalent systems may differ by more than just a unitary transformation. The construction itself will be trivial, and even showing the equivalence will not be hard, but we have to go through this in some detail as we could not find it published elsewhere.

\subsection{Doubling the dimension} \label{s21}
The new space $\mathcal{H}^B$ is simply constructed as the direct sum:
\begin{equation} \label{eVD1a}
\mathcal{H}^B \;=\; \mathcal{H}^A \oplus \mathcal{H}^A
\end{equation}
For finite-dimensional vector spaces, this step would double the dimension. Any vector of the new space $\mathcal{H}^B$ can be written as a pair $(\Psi, $$ \Phi)$ with the vectors $\Psi$ and $\Phi$ taken from $\mathcal{H}^A$. The inner product of two such pairs $(\Psi, $$ \Phi)$ and $(\Psi', $$ \Phi')$ is defined as the sum of two inner products from $\mathcal{H}^A$: 
\begin{equation} \label{eVD1b}
\Psi^\dag \Psi' + \Phi^\dag \Phi' 
\end{equation}
With this inner product, $\mathcal{H}^B$ also becomes a Hilbert space \cite[section 4.19]{dunford1958linear}. 

Any operator $M^A$ on the original space $\mathcal{H}^A$ can be lifted to an analogous operator $M^B$ on $\mathcal{H}^B$ via:
\begin{equation} \label{eVD2a}
M^B (\Psi, \Phi) \;=\; (M^A \Psi, M^A \Phi)
\end{equation}
for any $\Psi$ and $\Phi$ in $\mathcal{H}^A$. Clearly, this preserves all the algebraic relationships between operators. For example:
\begin{eqnarray} \nonumber 
R^A = M^A + N^A &\Rightarrow& R^B = M^B + N^B 
\\ \label{eVD2t} 
R^A = M^A N^A &\Rightarrow& R^B = M^B N^B 
\end{eqnarray}
for any operators $M^A$ and $N^A$, linear or otherwise, on $\mathcal{H}^A$. 

When $M^A$ is linear, $M^B$ is also linear, and their traces and adjoints can be lifted along the same lines. On finite Hilbert spaces, this is rather trivial, as traces and adjoints are then defined for any linear operator with full domain. When lifted to $\mathcal{H}^B$, the trace doubles:
\begin{equation} \label{eVD2tr}
\Tr{M^B} \;=\; 2 \Tr{M^A}
\end{equation} 
since it is calculated by summing over basis vectors and since \eq{eVD1a} doubles the size of the basis. It is also obvious that the adjoint can be lifted via:
\begin{eqnarray} \label{eVD2r}
R^A = (M^A)^\dag &\Rightarrow& R^B = (M^B)^\dag
\end{eqnarray}
since the inner product on $\mathcal{H}^B$, from \eq{eVD1b}, was based in a natural way on the inner product of $\mathcal{H}^A$. In particular, a self-adjoint operator stays self-adjoint, as it is lifted to $\mathcal{H}^B$, and a unitary operator stays unitary. On infinite Hilbert spaces, these relations will naturally hold only for operators whose trace or adjoint is defined at all. 

To construct system B, the original observables $O^A$ and the Hamiltonian $H^A$ are lifted, via \eq{eVD2a}, to $\mathcal{H}^B$ and become the new observables $O^B$ and Hamiltonian $H^B$:
\begin{eqnarray} \label{eVD2ao}
O^B (\Psi, \Phi) &=& (O^A \Psi, O^A \Phi) 
\\ \label{eVD2ah}
H^B (\Psi, \Phi) &=& (H^A \Psi, H^A \Phi) 
\end{eqnarray}
The original vacuum $\Theta^A$ gives rise to two states of lowest energy $(\Theta^A, 0)$ and $(0, \Theta^A)$ in system B, so that the new vacuum becomes twofold degenerate. In fact, every energy eigenstate becomes at least twofold degenerate.

Lifting density operators to $\mathcal{H}^B$ takes more care because of their normalization. Usually, density operators are required to have a trace of 1, but the lifting via \er{eVD2a} would double the trace. We thus define, for any density operator $\rho^A$ in system A, a corresponding density operator $\rho^B$ in system B as:
\begin{equation} \label{eVD2ar}
\rho^B (\Psi, \Phi) \;=\; \frac{1}{2} (\rho^A \Psi, \rho^A \Phi)
\end{equation}
The normalization factor $1/2$ ensures that $\rho^B$ has the same trace as $\rho^A$. We will see below that such normalization factors cancel so that the laws of quantum physics keep their usual shape in system B. Because $\rho^A$ is linear, self-adjoint, and positive semidefinite, it follows, from \eq{eVD2r} and \eq{eVD2ar}, that $\rho^B$ is also linear, self-adjoint, and positive semidefinite.

To lift a pure state, described by a vector $\Psi^A$, from $\mathcal{H}^A$ to $\mathcal{H}^B$, we can first turn it into a density operator $\rho^A = \Psi^A (\Psi^A)^\dag$ and then use \eq{eVD2ar}. This turns it into a mixed state $\rho^B$ of rank two. Alternatively, we could also identify the pure state in system A with another pure state in system B:
\begin{equation} \label{eVD2ap}
\Psi^B = (\Psi^A, 0)
\end{equation}
Section~\ref{s24} will show that these states $\rho^B$ and $\Psi^B$ are physically indistinguishable, in system B, so both can be identified with $\Psi^A$.

\subsection{Physical equivalence} \label{s22}
Since most relationships between operators are preserved, as we pass from system A to system B, the physical equivalence is not hard to show. Let us first consider the temporal evolution, the measurement process, and then, in section~\ref{s24}, the only non-trivial issue, the number of states.
  
The evolution of any state $\rho^A$ in system A, between quantum measurements, is given by the von Neumann equation: 
\begin{equation}  \label{eVP2a}
\frac{d}{dt}  \rho^A(t) \;=\; -i [H^A, \rho^A(t)]  
\end{equation}
(with $\hbar$ set to 1). Lifting $H^A$ and $\rho^A$, via \eq{eVD2ah} and \eq{eVD2ar}, and using the simple relations \er{eVD2t} gives:
\begin{equation} \label{eVP2b}
\frac{d}{dt}  \rho^B(t) \;=\; -i [H^B, \rho^B(t)]  
\end{equation}
as the normalization factor $1/2$, from \eq{eVD2ar}, occurs on both sides and cancels. Corresponding states $\rho^A$ and $\rho^B$ thus evolve in parallel. 

The results of measurements are also the same in systems A and B. For any observable $O^A$ of system A, the possible results are the eigenvalues $\lambda_n$ in the spectral expansion:
\begin{equation} 
O^A \;= \; \sum_n \lambda_n E^A_n
\end{equation} 
(where no two $\lambda_n$ are equal). Here the $E^A_n$ are orthogonal projections onto eigenspaces with:
\begin{equation} \nonumber
E^A_n E^A_m = \delta_{nm} E^A_n \; ; \;\; (E^A_n)^\dag = E^A_n
\end{equation}
for any $n$ and $m$. For simplicity, we have assumed that $O^A$ has discrete spectrum, as it would on a finite Hilbert space, but the generalization of the following to continuous spectra is straightforward. 

All the operators in this expansion can be lifted to the Hilbert space $\mathcal{H}^B$ via \eq{eVD2a}. Because of \eq{eVD2t} to \er{eVD2r}, these lifted operators $O^B$ and $E^B_n$ will satisfy the same relations as the original ones:
\begin{equation} \label{eVP2d}
O^B \;=\; \sum_n \lambda_n E^B_n
\end{equation} \begin{equation} \nonumber
E^B_n E^B_m = \delta_{nm} E^B_n \; ; \;\; (E^B_n)^\dag = E^B_n
\end{equation}
$O^B$ thus has the same eigenvalues as $O^A$. Even though $E^B_n$ has twice the rank of $E^A_n$, the possible results $\lambda_n$ of measurements are the same in systems A and B.

Both systems also agree in the probability of any particular result. In system A, the probability of measuring the result $\lambda_n$, in state $\rho^A$, for the observable $O^A$ is given by $\Tr{\rho^A E^A_n}$ (with $\Tr{\rho^A} = 1$). The analogous probability, in system B, is given by $\Tr{\rho^B E^B_n}$. Combining \eq{eVD2tr}, \er{eVD2ao}, and \er{eVD2ar} shows that both values match:
\begin{equation} \label{eVP4b}
\Tr{\rho^A E^A_n} \;=\; \Tr{\rho^B E^B_n}
\end{equation}
since the factor 2 from \eq{eVD2tr} cancels the factor $1/2$ from \eq{eVD2ar}. 

The same trace appears at the collapse of a wave function, where it keeps the trace of the density operators at 1:  
\begin{equation} \label{eVP4c}
\rho^A \to \frac{E^A_n \rho^A E^A_n}
{\Tr{\rho^A E^A_n}} \; ; \;\; 
\rho^B \to \frac{E^B_n \rho^B E^B_n}
{\Tr{\rho^B E^B_n}}
\end{equation}
According to \eq{eVP4b}, both traces are the same, and it follows that the relation \er{eVD2ar} between $\rho^A$ and $\rho^B$ continues to hold after a collapse of the wave function.

These results almost suffice to show the physical equivalence of the systems A and B. They guarantee that corresponding states $\rho^A$ and $\rho^B$, in both systems, evolve analogously and produce the same results during measurements. The only remaining issue is system B having more states than system A. We will resolve this issue, in section~\ref{s24}, by showing that certain states in system B are physically indistinguishable.

\subsection{The operators $V$ and $j$} \label{s23}
Before showing this, let us first introduce two linear operators $V$ and $j$ on $\mathcal{H}^B$. The former is defined as:
\begin{equation} \label{eVV1a}
V (\Psi,\Phi) \;=\; (\Phi, 0) 
\end{equation}
for any $\Psi$ and $\Phi$ from $\mathcal{H}^A$. Its adjoint satisfies:
\begin{equation} \label{eVV1b}
V^\dag (\Psi,\Phi) \;=\; (0, \Psi) 
\end{equation}
By using $V$ and $V^\dag$, we can switch between degenerate states. By definition, these operators have similar properties as fermionic field operators:
\begin{equation} \label{eVV2a}
V^2 \;=\; 0 \;=\; (V^\dag)^2 
\end{equation} \begin{equation} \label{eVV2b}
V V^\dag + V^\dag V \;=\; 1 
\end{equation}  
and it follows that:
\begin{equation} \label{eVV2d}
V V^\dag V \;=\; V (V^\dag V + V V^\dag) \;=\; V
\end{equation} \begin{equation} \label{eVV2e}
V^\dag V V^\dag \;=\; V^\dag (V V^\dag + V^\dag V) \;=\; V^\dag
\end{equation} 

The linear operator $j$, on $\mathcal{H}^B$, is defined as:
\begin{equation} \label{eVV3a}
j \;=\; V^\dag - V
\end{equation}
As already mentioned in \eq{eIN3}, $j$ satisfies:
\begin{eqnarray} \label{eVV3b}
j^2 &=& V^\dag (- V) + (- V) V^\dag \;=\; -1 
\\ \label{eVV3c}
j^\dag &=& V - V^\dag \;=\; -j
\end{eqnarray}
and is therefore unitary. It somewhat resembles an operator $\breve{J}$ used, instead of the imaginary unit $i$, on real Hilbert spaces \cite{stueckelberg1959field,stueckelberg1960quantum}. Unlike $\breve{J}$, our $j$ exists alongside $i$ and toggles between degenerate states. 

We can use $V$ to identify corresponding states in systems A and B. Let us first consider any linear operator $N^B$ that was constructed by lifting an operator $N^A$ from $\mathcal{H}^A$ via \eq{eVD2a}. According to \eq{eVV1a}, any such an operator $N^B$ commutes with $V$:
\begin{equation} \nonumber
V N^B (\Psi, \Phi)  \;=\; (N^A \Phi, 0)  \;=\; N^B V (\Psi, \Phi)
\end{equation}
Similarly, it commutes with $V^\dag$ and thus with $j$:
\begin{equation} \label{eVV4a}
[V, N^B] = 0; \; [V^\dag, N^B] = 0; \; [j, N^B] = 0
\end{equation}

On the other hand, any operator $M^B$, on $\mathcal{H}^B$, that fulfills the constraint \er{eVV4a} can be constructed by lifting an operator $M^A$ from $\mathcal{H}^A$. To show this, we choose $M^A$ so that for any $\Psi$ in $\mathcal{H}^A$:
\begin{equation} \label{eVV4b}
(M^A \Psi, 0) \;=\; V V^\dag M^B (\Psi, 0)
\end{equation}
Here the term $V V^\dag$, together with \eq{eVV1a} and \er{eVV1b}, guarantees that the second component of $(M^A \Psi, 0)$ is indeed zero. For any $\Phi$ in $\mathcal{H}^A$, we then get:
\begin{eqnarray} \nonumber
(M^A \Psi, 0) &=& V V^\dag M^B V V^\dag (\Psi, \Phi)
\\ \nonumber
(0, M^A \Phi) &=& V^\dag V V^\dag M^B V (\Psi, \Phi)
\end{eqnarray}
Because $M^B$ fulfills the constraint \er{eVV4a}, and because of \eq{eVV2d} and \er{eVV2e}, this becomes:
\begin{eqnarray} \nonumber
(M^A \Psi, 0) &=& V V^\dag M^B (\Psi, \Phi)
\\ \nonumber
(0, M^A \Phi) &=& V^\dag V M^B (\Psi, \Phi)
\end{eqnarray}
Summing both relations gives:
\begin{equation} \label{eVV4f}
(M^A \Psi, M^A \Phi) \;=\; M^B (\Psi, \Phi)
\end{equation}
which proves that $M^B$ can be constructed by lifting $M^A$ from $\mathcal{H}^A$ via \eq{eVD2a}.

\subsection{Indistinguishable states} \label{s24}
Using $V$ and $j$, we can now resolve the issue of system B having more possible states than system A. For any density operator $\rho^B_1$ on $\mathcal{H}^B$, we find another density operator $\rho^B_2$ on $\mathcal{H}^B$ that can be generated, via \eq{eVD2a}, by lifting a density operator $\rho^A$ from $\mathcal{H}^A$. We then show that all three density operators are physically equivalent, as they evolve in parallel and lead to the same results in quantum measurements, and conclude that the larger number of states, in system B, remains hidden.

For any given density operator $\rho^B_1$, on $\mathcal{H}^B$, the new $\rho^B_2$ is chosen as:
\begin{equation} \label{eVO1a}
\rho^B_2 \;=\; \frac{1}{4} \sum_{a,b = 0}^1 (V^\dag + V)^a j^b \rho^B_1
j^{-b} (V^\dag + V)^{-a} 
\end{equation}
This choice is motivated by $j$ and $(V^\dag + V)$ being unitary, due to \eq{eVV3b}, \er{eVV3c}, and: 
\begin{equation} \nonumber
(V^\dag + V)(V^\dag + V)^\dag \;=\; V^\dag V + V V^\dag \;=\; 1 
\end{equation}
With $\rho^B_1$ being self-adjoint and positive semidefinite, it follows that $\rho^B_2 $ is also self-adjoint and positive semidefinite. Due to the the cyclic property of the trace, we get:
\begin{equation} \label{eVO1t}
\;\;\; \Tr{(V^\dag + V)^a j^b \rho^B_1
j^{-b} (V^\dag + V)^{-a}} \;=\; \Tr{\rho^B_1}
\end{equation}
so that $\rho^B_2$ has the same trace as $\rho^B_1$ and qualifies as density operator.

Let us first show that $\rho^B_2$ commutes with $V$ and $V^\dag$. For any operator $R$ on $\mathcal{H}^B$, the sum:
\begin{equation} \nonumber
\sum_{a = 0}^1 (V^\dag + V)^a R (V^\dag + V)^{-a} 
\end{equation}
commutes with $(V^\dag + V)$ since $(V^\dag + V)^2 = 1$. Because \eq{eVO1a} is based on such a sum, $\rho^B_2 $ commutes with $(V^\dag + V)$. Similarly, we can use:
\begin{equation} \nonumber
(V^\dag + V) j \;=\; -V^\dag V + V V^\dag \;=\; - j (V^\dag + V)
\end{equation}
to move $j$ and $j^{-1}$ past $(V^\dag + V)$ in \eq{eVO1a}. From $j^2 = -1$, it then follows that $\rho^B_2 $ commutes with $j$.

Taken together, this proves that $\rho^B_2 $ commutes with $2V = V^\dag + V - j$ and with $2V^\dag = V^\dag + V + j$:
\begin{equation} \nonumber
[\rho^B_2, V] \;=\; 0 \;=\; [\rho^B_2, V^\dag]
\end{equation} 
and fulfills the constraint \er{eVV4a}. We can therefore find, via \eq{eVV4b}, an operator on $\mathcal{H}^A$ that becomes $\rho^B_2$ when lifted to $\mathcal{H}^B$. Calling this operator $\rho^A/2$, we get: 
\begin{equation} \label{eVO2f}
\rho^B_2 (\Psi, \Phi) \;=\; \frac{1}{2} (\rho^A \Psi, \rho^A \Phi)
\end{equation}
(for any $\Psi$ and $\Phi$ in $\mathcal{H}^A$). We already know, from section~\ref{s22}, that these states $\rho^A$ and $\rho^B_2$ are physically equivalent. 

All that remains to be shown is that $\rho^B_2$ is equivalent to $\rho^B_1$. To show this, we recall that all the observables $O^B$ of system B were lifted from $\mathcal{H}^A$. Because of \eq{eVV4a}, they commute with $V$, $V^\dag$, and $j$. This also holds for the Hamiltonian $H^B$ which thus commutes with all the factors surrounding $\rho^B_1$ in the definition \er{eVO1a} of $\rho^B_2$. This implies that the relation \er{eVO1a} remains valid as $\rho^B_1$ and $\rho^B_2$ evolve under the Hamiltonian $H^B$ in the von Neumann equation.

We also know, from \eq{eVP2d}, that the $E^B_n$, which project onto eigenspaces of observables, were lifted from $\mathcal{H}^A$ and also commute with $V$, $V^\dag$, and $j$. As the collapse of the wave function, in \eq{eVP4c}, is described by $E^B_n$, it follows that the relation \er{eVO1a}, between $\rho^B_1$ and $\rho^B_2$, remains valid during this collapse. 

Finally, the probabilities of measuring results $\lambda_n$, for any observable $O^B$, are also the same in state $\rho^B_1$ and state $\rho^B_2$. This follows from the analogue of \eq{eVO1t}
\begin{equation} \nonumber
\Tr{(V^\dag + V)^a j^b \rho^B_1 E^B_n 
j^{-b} (V^\dag + V)^{-a}} \;=\; \Tr{\rho^B_1 E^B_n}
\end{equation}
Because $E^B_n$ commutes with $(V^\dag + V)$ and $j$, we conclude from \eq{eVO1a} that:
\begin{equation}
\Tr{\rho^B_2 E^B_n} \;=\; \Tr{ \rho^B_1 E^B_n }
\end{equation}

Taken together, this shows that the three density operators $\rho^B_1$, $\rho^B_2$, and $\rho^A$ evolve in parallel and produce the same results during measurements. Any density operator $\rho^B_1$ in system B thus corresponds to a physically equivalent operator $\rho^A$ in system A, and vice versa, via a many-to-one relationship. Both systems are physically equivalent, and since $\rho^B_1$ and $\rho^B_2$ cannot be distinguished by observations, the larger number of states in system B remains hidden. In particular, the vacuum degeneracy of system B stays hidden.

We can also conclude that it does not matter whether a pure state $\Psi^A$ is lifted to $\mathcal{H}^B$ via \eq{eVD2ar} or \er{eVD2ap}. The former choice results in 
\begin{equation} \label{eVO9a}
\rho^B_2 \;=\; \frac{1}{2} \Psi^B_1 (\Psi^B_1)^\dag + \frac{1}{2} \Psi^B_2 (\Psi^B_2)^\dag
\end{equation}
with $\Psi^B_1 = (\Psi^A, 0)$ and $\Psi^B_2 = (0, \Psi^A)$. By contrast, \eq{eVD2ap} results in:
\begin{equation} 
\rho^B_1 \;=\; \Psi^B_1 (\Psi^B_1)^\dag 
\end{equation}
Inserting this $\rho^B_1$ into \eq{eVO1a}, and using $V^\dag \Psi^B_1 = \Psi^B_2$, reproduces the $\rho^B_2$ from \eq{eVO9a}. The two states are thus physically indistinguishable and can both be identified with $\Psi^A$.

\subsection{Physical arguments} \label{s25}
As this proof of physical equivalence was rather formal, let us briefly discuss it. Firstly, it should be noted that the degeneracy of states depends on the convention that every state is described by a density operator. Such states are sometimes called quantum ``microstates" in contrast to the  quantum ``macrostates" (not thermodynamic macrostates) that can actually be distinguished by observables \cite[section 11.5]{jauch1968foundations}. The two microstates 
$\rho^B_1$ and $\rho^B_2$, which cannot be distinguished, would belong to the same macrostate. One might thus argue, for example, that all vacuum states belong to the same macrostate and are not really degenerate, but this seems to be mostly a matter of terminology. No matter how states are defined, the two systems A and B will differ by more than a unitary transformation, but still be physically equivalent. 

Secondly, one might worry that doubling the number of microstates somehow violates the Pauli exclusion principle. It would be violated if we doubled, in atomic physics, the number of electrons in each orbital. However, this is not what we have done here. In a system with $n$ electrons, this doubling of orbitals could increase the number of possible states by as much as $2^n$. By contrast, in our system B, we have only doubled the number of possible states, no matter how many electrons the system contains. 

Finally, one might worry that doubling the number of states affects the sum over states in thermodynamics. The Gibbs formula for the entropy increases by $k_B \ln(2)$ when we double the number of states (assuming that corresponding states are assigned equal probability). This would be a problem if we could measure the absolute value of the Gibbs entropy, and not only the relative changes of entropy in the second law of thermodynamics.

For our systems A and B, this problem cannot arise. It is widely accepted that all of physics can be based on quantum physics, so that any measurement can, at least in principle, be regarded as a quantum measurement (e.g., \cite[chapter 11]{jauch1968foundations}). When we measure, for example, the pressure, volume, and temperature in a Carnot cycle, all these measurements could be described, in principle, as observing the positions of the dials of certain instruments, and such a measurement of position $Q$ can be put into the usual quantum-mechanical forms $\Tr{\rho Q}$ or $\Psi^\dag Q \Psi$. As long as these quantum measurements are the same, in systems A and B, we will arrive at the same physical conclusions, find the same thermodynamic laws, and observe the same entropy. In system A, or other systems with unique vacuum, this entropy will, as usually, agree with the Gibbs entropy formula. In system B, it may not agree, simply because the Gibbs entropy formula was not designed for systems with a hidden vacuum degeneracy. We can generalize this entropy formula to such systems as well, by subtracting a term $k_B \ln(2)$ for any twofold, hidden degeneracy of the vacuum, and will thereby get the same value as in system A. Subtracting this term may seem unconventional, but it is similar to dividing sums over states by $N!$ in systems with $N$ identical particles. 

In the end, the proof of sections \ref{s22} to \ref{s24} always guarantees that the twofold degeneracy is hidden. Other kinds of vacuum degeneracy are also known from other contexts, especially spontaneously broken symmetries (e.g., \cite[chapter 20]{peskin1995introduction}), and it is accepted that such a degeneracy may remain hidden.

\section{From system B to C} \label{s3}
Introducing vacuum degeneracy was the first step towards our goal of making room for antilinear terms in the Hamiltonian. In the second, less obvious, step, we will replace some occurrences of the imaginary unit $i$, in observables, by the linear operator $j$ from \eq {eVV3a}. 

We will not replace all occurrences of $i$, in the laws of quantum physics, which would be analogous to doing quantum physics on a real Hilbert space \cite{stueckelberg1960quantum}. A physical state is commonly identified with a unit ray in Hilbert space, that is, with a set of vectors differing only by a complex phase $\exp(i a)$. This interpretation will work here as well, before and after the replacement of $i$, so that this particular $\exp(i a)$ is not replaced by $\exp(j a)$. Furthermore, we will take care that the abstract Schr\"odinger equation and the von Neumann equation maintain their usual form with the factor $i$, not $j$. 

We will not insist that all commutation relations between observables, such as $[Q, P] = i$ for position $Q$ and momentum $P$, maintain their usual form but let $i$ replace $j$ in such relations. Nevertheless, the resulting quantum system will still be physically equivalent to the original one. This will happen partly because $j$ shares some properties with $i$, and partly because the expectation values $\Tr{\rho O}$ and $\Psi^\dag O \Psi$ in quantum measurements are always real so that $i$ cannot be measured directly.

\subsection{The operators $K$, $L$, and $U$} \label{s31}
To properly replace $i$ with $j$, we first need to consider three other operators $K$, $L$, and $U$. The first operator $K$ is well-known and simply takes the complex conjugate of the vector to its right $K \Psi \;=\; \Psi^* $. It is thus antilinear. On finite-dimensional, complex vector spaces, one can find $K \Psi $ simply by taking the complex conjugate of each component of $\Psi$. On  infinite-dimensional Hilbert spaces, $K$ can be defined by specifying a basis that stays invariant under $K$, but we do not need to review these details here, as we 
will only need a few basic properties of $K$. 

Like other operators, $K$ can be lifted from $\mathcal{H}^A$ to $\mathcal{H}^B$. We will use it mostly on $\mathcal{H}^B$ and, for simplicity, write it as $K$, not $K^B$. Like any operator lifted from $\mathcal{H}^A$, this $K$ obeys \eq{eVV4a}:
\begin{equation} 
[V, K] = 0; \; [V^\dag, K] = 0; \; [j, K] = 0
\end{equation}
Two other basic properties of $K$ are $K^2 = 1$ and:
\begin{eqnarray} \label{eJM1b}
\left\langle K \Psi , K \Phi \right\rangle &=& \left\langle \Psi , \Phi \right\rangle^*
\end{eqnarray}
where $\left\langle \Psi , \Phi \right\rangle $ is the inner product of $\Psi$ and $\Phi$ (which we identify with $\Psi^\dag \Phi$ so that it is linear in the second argument). That is, taking the conjugate of two vectors also takes the conjugate of their inner product. Consequently, $K$ maps any orthonormal basis of the Hilbert space, consisting of vectors $\Gamma_n$, into another orthonormal basis consisting of vectors $K \Gamma_n$: 
\begin{eqnarray} \label{eJM1d}
\left\langle K \Gamma_m , K \Gamma_n \right\rangle &=& \delta_{mn}
\end{eqnarray}
since the Kronecker delta $\delta_{mn}$ is a real number. The complex conjugation thus acts somewhat like a unitary transformation, yet being antilinear, it is not exactly unitary (but antiunitary). In particular, it does not keep the trace of a linear operator $B$ invariant (with $\Tr{KBK} = \Tr{B}^*$ due to \eq{eJM1b}). All this does not fix $K$ completely, as we might, for example, still choose whether $K$ commutes with position $Q$ or with momentum $P$ in the canonical commutation relation $[Q, P] = i$, but we do not need to specify this here, as any definition of $K$ with the above properties suffices for our purposes. 

The second operator $L$ can be regarded as a counterpart to $K$ in the same sense that our $j = V^\dag - V
$, from \eq{eVV3a}, is a counterpart to $i$. It is defined on the degenerate Hilbert space $\mathcal{H}^B$ as:
\begin{equation} \label{eJM2a}
L \;=\; V V^\dag - V^\dag V 
\end{equation}
with the $V$ from \eq{eVV1a}. It is self-adjoint, and, being linear, it commutes with $i$. It anticommutes with $j$:
\begin{equation} 
L j \;=\; -V^\dag - V \;=\; - j L
\end{equation}
due to \eq{eVV2d} to \er{eVV3a}. By contrast, $K$ anticommutes with $i$ and commutes with $j$. Like $K$, this $L$ squares to:
\begin{equation} \label{eJM2c}
L^2 \;=\; V V^\dag V V^\dag + V^\dag V V^\dag V \;=\; 1
\end{equation} 
and is thus unitary ($L^\dag = L = L^{-1}$). Since $K$ commutes with $V$ and $V^\dag$, it also commutes with $L$. All in all, the factors $i$, $j$, $K$, and $L$ all commute with each other, except for $iK = -Ki$ and $jL = -Lj$. They square either to $i^2 = -1 = j^2$ or to $K^2 = 1 = L^2$.

The third operator $U$ is defined as:
\begin{equation} \label{eAU1a}
U \;=\; \frac{1}{2}\left( 1 - ij + KL + ijKL \right)
\end{equation}
in terms of the $j$, $K$, and $L$ from \eq{eVV3a}, \er{eJM1b}, and \er{eJM2a}. Since $i$, $j$, and $L$ are linear, but $K$ antilinear, this $U$ is neither linear nor antilinear. Recall that the product of a linear and an antilinear operator is antilinear, so $KL$ is antilinear, and $U$ is a sum of linear and antilinear terms. In general, the sum $M = B + A$ of a linear operator $B$ and an antilinear operator $A$ is called ``real-linear" \cite{huhtanen2010real}, so $U$ is real-linear. Note that any real-linear operator commuting with $i$ is, by definition, linear. The properties of real-linear operators are summarized in the appendix, but, for now, we can deal with $U$ simply by using the explicit expression \er{eAU1a} in terms of linear and antilinear operators. 

To interpret $U$, we use that $\frac{1}{2}( 1 - ij)$ is an orthogonal projection because of $(ij)^\dag = ij$, from \eq{eVV3c}, and because of:
\begin{eqnarray}
\left( \frac{1}{2}( 1 - ij) \right)^2 &=& \frac{1}{4}\left( 1 - 2ij + i^2j^2 \right) \nonumber \\ \nonumber
&=& \frac{1}{2}( 1 - ij)
\end{eqnarray}
Since $[ij,KL]=0$, this projection commutes with $U$, so $U$ acts separately on two orthogonal complements of the Hilbert space, the range and the kernel of this projection, consisting of vectors $\Psi$ with $ij \Psi = -\Psi$ and $ij \Psi = +\Psi$, respectively. The first two terms of \eq{eAU1a}:
\begin{equation} \nonumber
\frac{1}{2}\left( 1 - ij \right)
\end{equation}
act on the range of this projection and keep it invariant. The other two terms of \eq{eAU1a}:
\begin{equation} \nonumber
\frac{1}{2}\left( 1 + ij \right) KL 
\end{equation}
act on the kernel of this projection and multiply it by $KL$. From \eq{eJM2c}, we know that $L$ is unitary, and from \eq{eJM1d} we know that $K$ maps any orthonormal basis into another orthonormal basis. It follows that the operators $KL$ and $U$, despite being real-linear, also map any orthonormal basis into another orthonormal basis. Writing the basis vectors as $\Gamma_n$, we thus get:
\begin{equation} \label{eAU1ga}
 \left\langle U\Gamma_m, U\Gamma_n \right\rangle \;=\; \delta_{mn}
\end{equation}
Like the complex conjugation $K$ from \eq{eJM1d}, this $U$ therefore resembles a unitary transformation, except for not being linear. In particular, it is bounded, and we will even find $U^\dag = U^{-1}$ in \eq{eAD8v}.

The same interpretation of $U$, together with $(KL)^2 = 1$, also tells us that $U$ squares to 1: 
\begin{eqnarray}
U^2 &=& \frac{1}{4} \left( 1 - ij \right)^2 + \frac{1}{4} \left( 1 + ij \right)^2 \left(KL \right)^2
\nonumber \\ \label{eAU1u}
&=& \frac{1}{2} ( 1 - ij ) + \frac{1}{2} ( 1 + ij ) \;=\; 1
\end{eqnarray}
It is thus invertible with $U^{-1} = U$.

\subsection{Interchanging $i$ and $j$} \label{s32a}
Using the real-linear $U$, from \eq{eAU1a}, we can now introduce the desired transformation that interchanges $i$ and $j$ and turns system B into system C. We simply transform any vector $\Psi$ and any operator $M$ of the Hilbert space $\mathcal{H}^B$ as:
\begin{equation} \label{eAU2a}
\Psi \;\to\; U \Psi \; ; \;\;
M \;\to\; \trfu{M}
\end{equation}
Since $U$ is bounded, any bounded $M$ turns into a bounded $\trfu{M}$. Since $U^2 = 1$, applying the transformation twice leads back to the original $M$. 

For the most part, we will apply this transformation only to linear operators that commute with $j$, for example, to the observables $O^B$ of system B. For any such operator and, more generally, for any operator $M$ that commutes with $ij$, the transformation \er{eAU2a} becomes:
\begin{eqnarray}
\trfu{M} &=& \frac{1}{4} (1 - ij)^2 M + \frac{1}{4} (1 + ij)^2 KL M KL 
\nonumber \\ \label{eDT1a} &=&
\frac{1-ij}{2} M + \frac{1+ij}{2} KL M KL 
\end{eqnarray}
where we have used $(1-ij)(1+ij) = 0$ in the first step. The product of two antilinear operators, like $K$ and $L M KL$, is linear, so any linear $M$ that commutes with $j$ transforms into a linear $\trfu{M}$. In particular, $i$ transforms into $j$ and vice versa:
\begin{eqnarray}
\trfu{i} &=& \frac{1-ij}{2} i + \frac{1+ij}{2} (-i) \;=\; j 
\\ \label{eDT1j}
\trfu{j} &=& \frac{1-ij}{2} j + \frac{1+ij}{2} (-j) \;=\; i  
\end{eqnarray}
while their product $ij$ commutes with $U$.

Occasionally, we will apply this transformation to antilinear operators that commute with $j$. For any such operator and, more generally, for any operator $N$ that anticommutes with $ij$, the transformation \er{eAU2a} becomes:
\begin{eqnarray}
\trfu{N} &=& \frac{1}{4} (1 - ij)^2 N KL + \frac{1}{4} (1 + ij)^2 KL N 
\nonumber \\ \label{eAU2e} &=&
\frac{1-ij}{2} N KL + \frac{1+ij}{2} KL N
\end{eqnarray}
where we have again used $(1-ij)(1+ij) = 0$. In particular, $K$ transforms into $L$ and vice versa:
\begin{eqnarray}
\trfu{K} &=& \frac{1-ij}{2} L + \frac{1+ij}{2} L \;=\; L 
\\ \label{eDT1l}
\trfu{L} &=& \frac{1-ij}{2} K + \frac{1+ij}{2} K \;=\; K  
\end{eqnarray}
while their product $KL$ commutes with $U$. As before, any real-linear $N$ that commutes with $j$ maps into a $\trfu{N}$ that commutes with $\trfu{j} = i$ and is linear. This already indicates how we are going to achieve our goal of finding room for antilinear terms $N$ in the Hamiltonian. We will turn them into linear terms $\trfu{N}$ that can be added to the new Hamiltonian.

\subsection{The adjoint $U^\dag$} \label{s32b}
A slight complication, when using a real-linear $U$ on the complex Hilbert space, comes from finding the adjoint $U^\dag$ of a real-linear operator. Recall that the usual definition for the adjoint $B^\dag$ of a linear operator $B$:
\begin{eqnarray} \label{eAD3a}
\left\langle B^\dag \Psi , \Phi \right\rangle &=& \left\langle \Psi , B \Phi \right\rangle 
\end{eqnarray}
cannot be used to define the adjoint of an antilinear operator since then the left-hand side would be linear in $\Phi$ and the right-hand side antilinear. Instead, the adjoint $A^\dag$ of an antilinear operator $A$, which is familiar from time-reversal, is the unique operator satisfying (e.g., \cite{sharma1988semilinear}):
\begin{eqnarray} \label{eAD3b}
\left\langle A^\dag \Psi , \Phi \right\rangle &=& \left\langle \Psi , A \Phi \right\rangle^*
\end{eqnarray}
for any vectors $\Psi $ and $\Phi $ in the complex Hilbert space. Due to the complex conjugation on the right-hand side, both sides are linear in $\Phi$ and $\Psi$, with the adjoint $A^\dag$ being an antilinear operator. Since \eq{eAD3a} and \er{eAD3b} differ, neither of them will, in general, hold for a sum of linear and antilinear operators, that is, for real-linear operators.

However, both linear and antilinear operators $M$ satisfy the relation:
\begin{eqnarray} \label{eAD4b}
\Re \left\langle M^\dag \Psi , \Phi \right\rangle &=& \Re \left\langle \Psi , M \Phi \right\rangle 
\end{eqnarray}
which results from taking the real part of \eq{eAD3a} or \er{eAD3b}. It is thus plausible to define the adjoint $M^\dag$ of real-linear operators $M$ so that it also fulfills this relation (see appendix). Unfortunately, this definition of the adjoint of real-linear operators is only sometimes \cite{huhtanen2010real}, but not always \cite{pian1986adjoint}, used in the mathematical literature. We will use it here because it has convenient properties for our purpose. It agrees with how the adjoint of linear and antilinear operators is defined in \eq{eAD3a} and \er{eAD3b}, and obeys similar rules as the adjoint of linear operators. The appendix shows: 
\begin{eqnarray} \label{eAD8w}
(M^\dag)^\dag &=& M \\  \label{eAD8x}
 (M + N)^\dag &=& M^\dag + N^\dag 
 \\ \label{eAD8y}
 (M N)^\dag &=& N^\dag M^\dag
\end{eqnarray}
for any real-linear operators $M$ and $N$ where these adjoints exist.

It should be noted that the last rule, the product rule, may no longer hold when $N^\dag$ is replaced by the adjoint $\Psi^\dag$ of a vector $\Psi$. This happens because $\Psi^\dag$ does not map the Hilbert space into itself, but into the field $\mathbb{C}$ of complex numbers. The same issue already occurs for antilinear operators. When we write the inner product as $\left\langle \Psi , \Phi \right\rangle =$$\Psi^\dag \Phi$, then \eq{eAD3b} tells us that $(M \Psi)^\dag \Phi$ is, in general, not the same as $\Psi^\dag M^\dag \Phi$. To deal with such expressions, we should use \eq{eAD4b} instead. Apart from this, we can use the adjoints of real-linear operators about as easily as the adjoints of linear operators.

With these relations, we can find the adjoint $U^\dag$ of the real-linear operator $U$ from \eq{eAU1a}. First, comparing \eq{eJM1b} and \er{eAD3b}, we find the familiar relation $K^\dag = K$. From \eq{eVV3c} and \er{eJM2a}, we also know $j^\dag = -j$ and $L^\dag = L$. Taking the adjoint of \eq{eAU1a} gives:
\begin{equation} 
U^\dag \;=\; \frac{1}{2}\left( 1 - ji + LK + LKji \right) \;=\; U \label{eAD8u}
\end{equation}
With $U^2=1$, from \eq{eAU1u}, this yields:
\begin{equation} \label{eAD8v}
U^\dag \;=\; U^{-1}
\end{equation}
The transformation $U$ thus keeps the norm of any vector $\Psi$ invariant:
\begin{eqnarray}
|| U \Psi ||^2 &=& \Re \left\langle U \Psi , U \Psi \right\rangle 
\nonumber \\
&=& \Re \left\langle U^\dag U \Psi , \Psi \right\rangle
\nonumber \\ \label{eAD8r}
&=& || \Psi ||^2
\end{eqnarray}
where \eq{eAD4b} was used. Due to \eq{eAD8y}, any real-linear operator $M$ also obeys:
\begin{equation} \label{eAD8s}
\left( \trfu{M} \right)^\dag \;=\; \trfu{M^\dag}
\end{equation}
In most of the following applications, a linear $M$ will be turned into a $\trfu{M}$ that is also linear. In such cases, \eq{eAD8s} implies that a self-adjoint $M$ is turned into a self-adjoint $\trfu{M}$, and a unitary $M$ is turned into a unitary $\trfu{M}$. Incidentally, as $U$ was defined, in \eq{eAU1a}, in terms of linear and antilinear operators, we could prove these relations by using only the familiar adjoints of linear and antilinear operators, but it is useful to know that they hold for the adjoint of real-linear operators as well.

\subsection{The new observables $O^C$} \label{s33a}
Using this transformation $ U $, we now construct our system C where antilinear terms become linear and find a place in the Hamiltonian. Recall that, in section~\ref{s21}, we started with system A, with Hamiltonian $H^A$, observables $O^A$, and Hilbert space $\mathcal{H}^A$, and constructed an equivalent system B with Hamiltonian $H^B$, observables $O^B$, and Hilbert space $\mathcal{H}^B$. Any density operator $\rho^A$ in system A corresponded to a density operator $\rho^B$ in system B. 

We now take this one step further and apply the transformation $U$, from \eq{eAU2a}, to any observable $O^B$. The resulting observable, for system C, is:
\begin{equation} \label{eES1a}
O^C \;=\; U O^B U^{-1}
\end{equation}
Since all the observables $O^B$ of the earlier system were lifted from $\mathcal{H}^A$, they commute with $V$ and $V^\dag$ (\eq{eVV4a}) and therefore with the operators $j$ and $L$ from \eq{eVV3a} and \er{eJM2a}. Being linear, $O^B$ also commutes with $i$. Consequently, $O^C$ commutes with $U j U^{-1} = i$ and with $U i U^{-1} = j$. Any real-linear operator commuting with $i$ is linear, so $O^C$ is linear and, due to \eq{eAD8s}, also self-adjoint. Instead of \eq{eES1a}, we can also use \eq{eDT1a} to express $O^C$. It simplifies to:
\begin{equation} \label{eES1b}
O^C \;=\; \frac{1-ij}{2} O^B + \frac{1+ij}{2} K O^B K 
\end{equation}
because of $[O^B,L]=0$, $[K,L]=0$, and $L^2=1$. 

Many relations between the observables of system B remain valid in system C. For example, it follows directly from \eq{eES1a} and \er{eES1b} that adding or multiplying two observables, or multiplying them with a real number, gives analogous results in both systems B and C. Multiplying an observable with the imaginary unit $i$ would not give analogous results, since $U$ does not commute with $i$, but this is of no concern here as it would not give a self-adjoint observable either.

Incidentally, we could also split an observable $O^B$ into a real and an imaginary part: 
\begin{equation} \nonumber
\Re O^B = \frac{1}{2} (O^B + K O^B K) \; ; \;\; \Im O^B = \frac{1}{2i} (O^B - K O^B K)
\end{equation}
with $O^B = \Re O^B + i \Im O^B$. The new observable from \eq{eES1b} would turn out to be:
\begin{equation}
O^C \;=\; \Re O^B + j \Im O^B 
\end{equation}
which illustrates how $i$ is replaced by $j$.

The same transformation $U$ relates states $\Psi^B$ in system B to corresponding states $\Psi^C$ in system C:
\begin{equation} \label{eES1p}
\Psi^C \;=\; U \Psi^B
\end{equation}
For mixed states, described by density operators, the rules of correspondence are more complex and will be derived in section~\ref{s35}.

As $U$ is so similar to a unitary transformation, it is straightforward to see that the new observables $O^C$ produce the same results as the previous observables $O^B$. To simplify notation, let us consider observables with discrete spectrum, although the generalization to continuous spectrum is straightforward. In \eq{eVP2d}, we have already used the spectral expansion $O^B = \sum_n \lambda_n E^B_n$ of an observable in system B. The eigenvalues $\lambda_n$ are real and the $E^B_n$ are orthogonal projections onto eigenspaces. Since $O^B$ commutes with $j$, these $E^B_n$ also commute with $j$. Applying $U$, we find the analogous expansion:
\begin{equation} \label{eES4c}
O^C \;=\; \sum_n \lambda_n E^C_n
\end{equation}
where the eigenvalues $\lambda_n$ remain the same and the operators $E^C_n = \trfu{E^B_n}$ retain the properties from \eq{eVP2d}:
\begin{equation} \label{eES4d}
E^C_n E^C_m = \delta_{nm} E^C_n \; ; \;\; (E^C_n)^\dag = E^C_n
\end{equation}
due to \eq{eAD8s}. As $E^B_n$ commutes with $j$, $E^C_n$ commutes with $\trfu{j} = i$ and is not only real-linear, but linear. It is thus an orthogonal projection, and \eq{eES4c} describes the spectrum of $O^C$. 

When $O^B$ is observed in experiments, the possible results are the values $\lambda_n$. According to \eq{eES4c}, the same results are observed when measuring $O^C$. The probability of seeing each result, in state $\Psi^B$ or state $\Psi^C$, is also the same:
\begin{eqnarray} 
(\Psi^C)^\dag E^C_n \Psi^C &=& 
\Re \left\langle \Psi^C, E^C_n \Psi^C \right\rangle
\nonumber \\  &=& 
\Re \left\langle U \Psi^B, U E^B_n \Psi^B \right\rangle
\nonumber \\  &=& 
\Re \left\langle U^\dag U \Psi^B, E^B_n \Psi^B \right\rangle
\nonumber \\  &=& \label{eES7e}
(\Psi^B)^\dag E^B_n \Psi^B 
\end{eqnarray}
where we have used \eq{eAD4b}, \er{eAD8v}, and the fact that the expectation value of a self-adjoint operator is real. In particular, the mean observed value of the observable is the same in both systems:
\begin{eqnarray} 
(\Psi^C)^\dag O^C \Psi^C &=& \sum_n \lambda_n (\Psi^C)^\dag E^C_n \Psi^C
\nonumber \\ &=&
(\Psi^B)^\dag O^B \Psi^B  \label{eES7f}
\end{eqnarray}
This shows that the transformation $U$, despite not being linear, does not affect the results of observations.

Both systems also reach corresponding states after a collapse of the wavefunction. Apart from a normalization factor, the state $\Psi^B$ becomes $E^B_n \Psi^B$ and the state $\Psi^C$ becomes $E^C_n \Psi^C$. The relationship $\Psi^C = U \Psi^B$ thus continues to hold after a collapse of the wave function:
\begin{equation} \label{eES7g}
E^C_n \Psi^C \;=\; U E^B_n U^{-1} U \Psi^B \;=\; U E^B_n \Psi^B
\end{equation}
and the norms of these states also stay equal due to \eq{eAD8r}. Section~\ref{s35} will show that this also holds for mixed states, so that corresponding observations in both quantum systems give the same results.

\subsection{The new Hamiltonian $H^C$} \label{s33b}
To show the physical equivalence of both systems, we still have to show that corresponding states evolve in parallel. When $\Psi^C(0) = U \Psi^B(0)$ holds at time 0, then it should continue to hold at any later time:
\begin{equation} \label{eEQ1x}
\Psi^C(t) = U \Psi^B(t)
\end{equation}
The main issue here is the form of the abstract Schr\"odinger equation. In system B it reads, as usually:
\begin{equation} \label{eEQ8a}
\frac{d}{dt} \Psi^B(t) \;=\; -i H^B \Psi^B(t) 
\end{equation}
After the transformation $U$, this becomes:
\begin{equation} \label{eEQ8b}
\frac{d}{dt} \Psi^C(t) \;=\; -j \trfu{H^B} \Psi^C(t) 
\end{equation}
where $\trfu{H^B}$ is the observable of energy, in system C, and $j$ replaces $i$ due to $\trfu{i} = j$. By construction, \eq{eEQ8b} guarantees that relation \er{eEQ1x} continues to hold as $\Psi^C(t)$ evolves. 

However, the substitution of $j$, at such a central place of quantum physics, seems awkward. To avoid it, and retain the usual form of the Schr\"odinger equation, we take another step that seems somewhat less awkward. We let the Hamiltonian of system C, which we write as $H^C$, differ slightly from the observable of energy which we continue to write as $\trfu{H^B}$. Specifically, we set the Hamiltonian to:
\begin{equation} \label{eEQ3b}
H^C \;=\; -ij \, \trfu{H^B}
\end{equation}
so that \eq{eEQ8b} becomes:
\begin{equation} \label{eEQ8d}
\frac{d}{dt} \Psi^C(t) \;=\; -i H^C \Psi^C(t) 
\end{equation}
with the usual factor $i$ instead of $j$. 

Like any observable $O^C$ from \eq{eES1a}, the observable $\trfu{H^B}$ is linear, self-adjoint, and commutes with $j$. With $j^\dag = -j$, it follows that the Hamiltonian $H^C$ is also linear, commutes with $j$, and is self-adjoint:
\begin{eqnarray} 
\left( H^C \right)^\dag &=& -\trfu{H^B} (-j) (-i) \;=\; H^C 
\end{eqnarray}
as it should be to guarantee the condition of unitarity. Moreover, it follows that $\trfu{H^B}$ commutes with $H^C$ so that the energy $(\Psi^C)^\dag \trfu{H^B} \Psi^C$ is conserved. Section~\ref{s35} will show that the same Hamiltonian $H^C$ also appears in the von Neumann equation of system C. 

Making this distinction, between the Hamiltonian $H^C$ and the observable $\trfu{H^B}$ of energy, is unconventional. Even when non-Hermitian Hamiltonians are used in quantum mechanics \cite{bender2002complex}, it is commonly assumed that the Hamiltonian should be equal to the observable of energy and thus have only real eigenvalues since energies are real-valued. In our case, both $H^C$ and $\trfu{H^B}$ have real eigenvalues, since they are self-adjoint, but only those of $\trfu{H^B}$ denote energy values. 

Their two spectra are, however, closely related. The factor $-ij$, from \eq{eEQ3b}, has the properties 
\begin{equation} \nonumber
(-ij)^2 = 1 \; ; \;\;  (-ij)^\dag (-ij) = 1
\end{equation}
which makes it drop out of many calculations. It is self-adjoint and commutes with the self-adjoint operator $\trfu{H^B}$, so both operators can be diagonalized simultaneously. Since $-ij$ squares to 1, it constitutes a ``grading" operator whose eigenvalues are either $+1$ or $-1$, and the eigenvalues of the Hamiltonian $H^C$ can differ from those of the observable $\trfu{H^B}$ by at most a sign. 

In fact, making a distinction between the Hamiltonian and the observable of energy is not without precedent. As a trivial example, consider an experiment where we measure all the energy within a box except for the energy of neutrinos passing, without interaction, through this box. The observable being measured in this experiment will contain no contribution from neutrinos, but the Hamiltonian will still contain the terms describing neutrino propagation. More generally, such a distinction between Hamiltonian and observable of energy is likely to crop up whenever the Hamiltonian describes some process, involving neutrinos, dark matter, or perhaps degenerate vacuum states, that we cannot observe directly. 

As another example, consider the Hamiltonian in gauge theories. The observable of momentum $m \dot{Q}$ will, in general, differ from the generator $P$ of spatial displacements, by more than just a factor $\hbar$, because the observable $m \dot{Q}$ (or at least its expectation value) stays invariant under gauge transformations while the generator $P$ does not \cite[section 13.5]{jauch1968foundations}. Presumably the same also applies, for certain choices of the gauge, to the temporal dimension, where displacements are generated by the Hamiltonian. If we chose a gauge that varies in time, and derived the Hamiltonian, it would presumably also vary in time, even if the observable of energy did not. Our distinction between the Hamiltonian and the observable of energy is not exactly the same as in these simple examples, but it also seems permissible as long as it produces the correct physical predictions.

The distinction between $H^C$ and $\trfu{H^B}$ is arguably the most unconventional feature of system C, but there are other ones. In many relations between operators, where $i$ appears explicitly, it will be replaced by $j$. The canonical commutation relation $[Q^B,P^B] = i$ between position $Q^B$ and momentum $P^B$ of a particle in one dimension, for example, will become:
\begin{equation} \label{eEQ6a}
 [Q^C, P^C] =  U [Q^B, P^B] U^{-1} = \trfu{i} = j 
\end{equation}
It has been argued that such commutation relations always need a term like $i$ on the right-hand side so that the Heisenberg uncertainty relation holds \cite{stueckelberg1960quantum}, but our $j$ is similar enough to $i$ to meet this requirement. Physical equivalence guarantees that the standard deviations of $Q^C$ and $P^C$, which can be observed, keep their usual values and satisfy the uncertainty relation.

\subsection{Mixed states} \label{s35}
So far, the physical equivalence of the two systems B and C has been proven only for pure states. Section~\ref{s33b} has shown that two corresponding states $\Psi^B(t)$ and $\Psi^C(t)$ evolve in parallel, between measurements, and section~\ref{s33a} has shown that they produce the same results when corresponding observables $O^B$ and $O^C$ are measured. To finish this proof, let us now show the same results for mixed states.

As in section~\ref{s21}, the density operators $\rho^B$ or $\rho^C$ are constrained only by the usual requirements. They have to be linear, self-adjoint, and positive semidefinite with trace 1. Unlike the observables $O^B$ or $O^C$, they do not have to commute with $j$. This prevents us from simply applying the transformation $U$ to find corresponding density operators. When $\rho^B$ does not commute with $j$, then $\trfu{\rho^B}$ does not commute with $\trfu{j}=i$ and is not linear. 

To find the correct relation between $\rho^B$ and $\rho^C$, let us first consider a density operator $\rho^B$ of finite rank. With the spectral theorem of self-adjoint operators, it can be written as:
\begin{equation} \label{eMS1a}
\rho^B \;=\; \sum_n p_n \Psi^B_n (\Psi^B_n)^\dag
\end{equation}
where the non-negative coefficients $p_n$ add up to 1 and the vectors $\Psi^B_n$ are orthogonal to each other and normalized to $||\Psi^B_n|| = 1$. As usually, we can interpret this as a statistical mixture of pure states $\Psi^B_n$ occurring with probability $p_n$. From \eq{eES1p}, we know, for each of these states $\Psi^B_n$ in system B, the corresponding state $\Psi^C_n$ in system C:
\begin{equation} \label{eMS1b}
\Psi^C_n \;=\; U \Psi^B_n
\end{equation}
This implies that the corresponding density operator $\rho^C$, in system C, is given by:
\begin{equation} \label{eMS1c}
\rho^C \;=\; \sum_n p_n (U \Psi^B_n) (U \Psi^B_n)^\dag
\end{equation}
We have seen, in section~\ref{s32b}, that $(U \Psi^B_n)^\dag$ is not necessarily the same as $(\Psi^B_n)^\dag U^\dag$, since $U$ is not linear, so the expression \er{eMS1c} is not the same as $U \rho^B U^\dag$. 

Unlike $U \rho^B U^\dag$, this $\rho^C$ is always linear since $(U \Psi^B_n)$ is just another vector without any antilinear or real-linear properties. The vectors $U \Psi^B_n$ are still normalized to 1 and orthogonal to each other, as \eq{eAU1ga} tells us that $U$ maps any orthonormal basis into another orthonormal basis. Consequently, $\rho^C$ can be interpreted, like $\rho^B$, as a statistical mixture of pure states with probability $p_n$. Like $\rho^B$, the $\rho^C$ from \eq{eMS1c} is self-adjoint and positive semidefinite with trace $\sum_n p_n = 1$.

The equivalence of $\rho^B$ and $\rho^C$, within their respective quantum systems, follows from the equivalence of the pure states $\Psi^B_n$ and $\Psi^C_n = U \Psi^B_n$. From section~\ref{s33a}, we know that measuring an observable $O^B$, in state $\Psi^B_n$, and the corresponding observable $O^C$, in state $\Psi^C_n$, produces the same results. This easily generalizes to density operators. In particular, it follows from:
\begin{eqnarray} \nonumber
\Tr{\rho^B O^B} &=&  \sum_n p_n (\Psi^B_n)^\dag O^B \Psi^B_n
\\  \nonumber
\Tr{\rho^C O^C} &=&  \sum_n p_n (\Psi^C_n)^\dag O^C \Psi^C_n
\end{eqnarray}
and the earlier result \er{eES7f} that the expectation values of the observables are the same:
\begin{equation} 
\Tr{\rho^B O^B} \;=\; \Tr{\rho^C O^C}
\end{equation}
The same equality holds, due to \eq{eES7e}, for the probabilities $\Tr{\rho^B E^B_n}$ or $\Tr{\rho^C E^C_n}$ of observing any particular eigenvalue of $O^B$ or $O^C$. 

Similarly, the results of section~\ref{s33b} can be used to show that the density matrices $\rho^B$ and $\rho^C$ evolve in parallel. We find their evolution by applying the abstract Schr\"odinger equation, from \eq{eEQ8a} and \er{eEQ8d}, to the vectors $\Psi^B_n$ and $\Psi^C_n$ from \eq{eMS1a} to \er{eMS1c}. This gives the von Neumann equation:
\begin{eqnarray}  \label{eMS3a}
\frac{d}{dt}  \rho^B(t) &=& -i [H^B, \rho^B(t)]  \\
\frac{d}{dt}  \rho^C(t) &=& -i [H^C, \rho^C(t)]  
\end{eqnarray}
where the Hamiltonian $H^C$, in system C, is again given by the $ -ij \, \trfu{H^B}$ from \eq{eEQ3b}. Here we have used: 
\begin{equation} \nonumber
(H^C \Psi^C_n)^\dag \;=\; (\Psi^C_n)^\dag H^C 
\end{equation}
which holds trivially since $H^C$ is linear and self-adjoint. Note that the von Neumann equation keeps its usual form with a factor $i$, not $j$. In fact, if we had not already included the factor $-ij$ in the Hamiltonian $H^C$, in \eq{eEQ3b}, we would have to be careful where to put it now since $j$ does not necessarily commute with $\rho^C$. It is known that other placements of $j$, within the context of quantum physics on real Hilbert spaces, may lead to difficulties \cite{myrheim1999quantum}.

Both density matrices also continue to evolve in parallel after a collapse of the wave function. Again, we can show this by applying the corresponding result for pure states, from \eq{eES7g}, to the density matrices in \eq{eMS1a} and \er{eMS1c}. Apart from a trivial prefactor, which keeps the trace at 1, this gives:
\begin{eqnarray} 
\rho^B &\to& E^B_m \rho^B E^B_m \\
\rho^C &\to& E^C_m \rho^C E^C_m 
\end{eqnarray}
where we have used that $E^C_m$, from \eq{eES4d}, is linear and self-adjoint with $(E^C_m \Psi^C_n)^\dag=$$(\Psi^C_n)^\dag E^C_m$. The collapse thus takes the same familiar form in systems B and C.

These results can be generalized to density matrices that are not of finite rank. They clearly still hold when the sum over eigenstates in \eq{eMS1a} is infinite, and it is straightforward to generalize them to a continuous spectrum as well. In fact, there is another, more general way to show the same results. We could rewrite the linear density operator $\rho^B$, from \eq{eMS1a}, in terms of a real-linear operator $\rho^B_R$ with:
\begin{eqnarray}
\rho^B &=& \rho^B_R - i \rho^B_R i  \\
\rho^B_R \Phi &=& \sum_n p_n \Psi^B_n \Re (\Psi^B_n)^\dag \Phi
\end{eqnarray}
for any vector $\Phi$ in $\mathcal{H}^B$. Here $\Re$ acts on the whole subsequent product $(\Psi^B_n)^\dag \Phi$ and not just the first factor. This decomposition is analogous to the more familiar \eq{eSC5b} in the appendix. After putting $\rho^B$ into this form, we could then use \eq{eSC6z} to write $\rho^C$ from \eq{eMS1c} as: 
\begin{equation}
\rho^C \;=\; U \rho^B_R U^\dag - i U \rho^B_R U^\dag i 
\end{equation}
and conclude that $\rho^B_R$ simply becomes $U \rho^B_R U^\dag$ in system C, even though $\rho^B$ itself does not transform in such a simple way. Even without using the physical equivalence of pure states, it would then be straightforward to prove the physical equivalence of these density operators (not shown).

\section{Applications} \label{s5}
\subsection{Finding room for antilinear terms in the Hamiltonian} \label{s51}
This physical equivalence of the quantum systems A, B, and C, shown in the previous sections, may be useful for several applications involving antilinear operators. Our main goal, from the introduction, was to find room for antilinear terms in the Hamiltonian. That is, we would like to take a quantum system, with the usual, linear Hamiltonian $H^A$, and add an antilinear term $H^A_2$ so that states evolve as:
\begin{equation} \label{eAH0a}
\frac{d}{dt} \Psi^A(t) \;=\; -i (H^A + H^A_2) \Psi^A(t) 
\end{equation}
While this makes sense as a differential equation, it does not make sense as a Schr\"odinger equation because the Hamiltonian would not be linear. 

However, we can pass from system A to the system C, replace $H^A$ by the Hamiltonian $H^C$ from \eq{eEQ3b}, replace $H^A_2$ by an analogous term:
\begin{equation} \label{eAH1a}
H^C_2 \;=\; -ij U H^B_2 U^{-1}
\end{equation}
and replace $\Psi^A$ by $\Psi^C = U (\Psi^A, 0)$ according to \eq{eVD2ap} and \er{eES1p}. Lifting \eq{eAH0a} to the new Hilbert space $\mathcal{H}^B$, and applying the transformation $U$, gives:
\begin{eqnarray}
\frac{d}{dt} \Psi^C(t) &=& -U i (H^B + H^B_2) U^{-1} \Psi^C(t) 
\nonumber \\ 
&=& -i (H^C + H^C_2) \Psi^C(t) 
\end{eqnarray}
as in \eq{eEQ1x} to \er{eEQ8d}. In this equivalent form, the differential equation can be interpreted as a Schr\"odinger equation. Like other operators lifted to $\mathcal{H}^B$, the $H^B_2$ obeys \eq{eVV4a} and commutes with $j$. Consequently, $U H^B_2 U^{-1}$ commutes with $U j U^{-1} = i$ and is linear, so $H^C_2$ is also linear. In fact, we could start with any real-linear $H^A_2$, not just antilinear ones, and $H^C_2$ would still be linear. Though $H^A_2$ cannot be added directly to the Hamiltonian of system A, we can thus construct an equivalent system C where the corresponding term $H^C_2$ can be added.

Not every antilinear term $H^A_2$ can be added in this way. The main restriction is that the resulting Hamiltonian $H^C + H^C_2$ should still be self-adjoint, as required by the condition of unitarity. Because $H^C$ is self-adjoint, $H^C_2$ has to be self-adjoint. This condition is satisfied by any $H^A_2$ with:
\begin{equation} \label{eAH1f}
(i H^A_2)^\dag \;=\; -i H^A_2
\end{equation}
as we then get $-i H^B_2 = (i H^B_2)^\dag$. Applying $U$ yields:
\begin{equation} \nonumber
\left(j \trfu{H^B_2}\right)^\dag \;=\; -j \trfu{H^B_2}
\end{equation}
due to \eq{eAD8v} and $\trfu{i} = j$. Since $U H^B_2 U^{-1}$ is linear, we can conclude from \eq{eAH1a} that:
\begin{equation}
H^C_2 \;=\; (H^C_2)^\dag 
\end{equation}
By reversing this argument, we can also show that condition \er{eAH1f} is necessary for $H^C_2$ to be self-adjoint. 

Incidentally, if we let vectors evolve directly under \eq{eAH0a}, their norm stays constant, due to:
\begin{equation} \nonumber
\exp(-i H^A t - i H^A_2 t)^\dag \;=\; \exp(i H^A t + i H^A_2 t)
\end{equation}
and \eq{eAD4b}. By contrast, the inner product of two distinct vectors does not necessarily stay constant since it is, in general, not real. This illustrates the underlying reason why an antilinear $H^A_2$ cannot be added directly in system A. It might be possible to find a way around this issue, and add $H^A_2$ directly to $H^A$, but this would probably require that we change the laws of quantum physics substantially. We may have to treat two vectors $\Psi$ and $i \Psi$, differing only by a phase $i$, as distinct, yet indistinguishable, states instead of the same physical state. By passing from system A to system C, we avoid this tricky issue. As the Hamiltonian $H^C + H^C_2$ is linear and self-adjoint, two vectors $\Psi$ and $i \Psi$ can, as usually, be regarded as belonging to the same physical state, and the inner product of any two vectors will stay constant while they evolve. 

It should be acknowledged that adding the new term to the Hamiltonian in \eq{eAH1a} can change the physical properties of system C substantially so that the vacuum degeneracy may no longer be hidden and observables may take other forms. In particular, the subtle distinction between the observable of energy and the Hamiltonian, from \eq{eEQ3b}, might vanish. The precise form of observables depends, however, on the details of the quantum system and cannot be explored here.

\subsection{Linear time-reversal}  \label{s52a}
As another application, consider the case where system A has time-reversal symmetry $T^A$. Usually, this $T^A$ is an antilinear operator and there are good reasons for this \cite{wigner1931}. For example, when describing a particle with position $Q^A$ and momentum $P^A$, we would like $P^A$ to reverse under $T^A$ and $Q^A$ to stay invariant. The canonical commutation relation $[Q^A, P^A] = i$ then requires that $T^A$ anticommutes with $i$. More generally, $T^A$ should commute with the observable of energy:
\begin{equation} \label{TC1a}
[T^A, H^A] \;=\; 0
\end{equation}
so that it keeps energies invariant. It should also anticommute with the term $iH^A$ in the abstract Schr\"odinger equation so that it can reverse time. Again, this forces $T^A$ to be antilinear, and due to Wigner's theorem, even antiunitary \cite{wigner1931}:
\begin{equation} \label{TC1b}
(T^A)^\dag \;=\; (T^A)^{-1}
\end{equation}

Interestingly, neither of these arguments holds in system C. As it is physically equivalent to system A, it should also have a time-reversal operator. We can find this $T^C$ in analogy to the observables $O^C$ from section~\ref{s33a}. We lift $T^A$ to $\mathcal{H}^B$, where it becomes $T^B$, and then set:
\begin{equation}
T^C \;=\; U T^B U^{-1}
\end{equation}
Just like $O^C$, this $T^C$ turns out to be linear. Since $T^A$ anticommutes with $i$, its lifted version $T^B$ also anticommutes with $i$ but commutes, like other lifted operators, with $j$. This implies that $T^C$ anticommutes with $U i U^{-1} = j$ but commutes with $U j U^{-1} = i$ and is therefore linear. From \eq{TC1b}, we get $(T^B)^\dag = (T^B)^{-1}$ and, with \eq{eAD8v}:
\begin{equation}
(T^C)^\dag \;=\; (T^C)^{-1}
\end{equation}
so that $T^C$ is not only linear but unitary. 

Furthermore, it follows from \eq{TC1a} and the other properties of $T^A$ that $T^C$ commutes with the observable $U H^B U^{-1}$ of energy and the observable $Q^C = U Q^B U^{-1}$ of position, but anticommutes with the observable $P^C = U P^B U^{-1}$ of momentum, just as a time-reversal operator should. $T^C$ thus anticommutes with the product $Q^C P^C$, but this does not prevent it from being linear, since we know from \eq{eEQ6a} that the canonical commutation relation $[Q^C, P^C] = j$ now contains $j$, not $i$, on the right-hand side. Despite being linear, $T^C$ can thus anticommute with both sides of this rule.

Similarly, $T^C$ can reverse time in the Schr\"odinger equation despite being linear and commuting with the observable $U H^B U^{-1}$ of energy. This is only possible because of the unconventional distinction between the observable $U H^B U^{-1}$ and the Hamiltonian $H^C$ from \eq{eEQ3b}. As $T^C$ commutes with $i$, anticommutes with $j$, and commutes with $U H^B U^{-1}$, it anticommutes with the Hamiltonian:
\begin{equation}
\{ T^C, H^C \} \;=\; 
\{ T^C, -ij \, \trfu{H^B} \} \;=\; 0
\end{equation}
and thus anticommutes with the term $i H^C$ in the Schr\"odinger equation \er{eEQ8d}. If $\Psi^C(t)$ is a solution of that equation, then $T^C \Psi^C(t)$ solves the time-reversed equation:
\begin{eqnarray} 
\frac{d}{dt} T^C \Psi^C(t) &=& +i H^C T^C \Psi^C(t) 
\end{eqnarray}
While system C is mathematically more complicated than system A in some respects, the hidden degeneracy and the substitution of $j$ for $i$, it thus has a linear time-reversal operator $T^C$ and is simpler in this respect. It would be interesting to explore whether such a $T^C$, or its generalization to CPT, can be embedded in a continuous set of linear symmetries.

\subsection{Continuous symmetries} \label{s52b}
A similar argument also holds for generators of continuous symmetries. Usually, such generators $G$ have to be linear, so that the symmetry $1 + i \epsilon G$, for infinitesimal $\epsilon$, is linear and abides by Wigner's unitary-antiunitary theorem \cite{wigner1931}. However, a real-linear operator in system A corresponds to a linear operator in system C, and it thus makes sense to consider real-linear generators $G^A$  corresponding to linear $G^C$. To keep the norm constant, such a continuous symmetry would have to obey:
\begin{equation} 
(1 + i \epsilon G^A)^\dag \;=\; (1 + i \epsilon G^A)^{-1}
\end{equation}
Its generator would therefore be constrained by $(i G^A)^\dag = -i G^A$, like the $H^A_2$ from \eq{eAH1f}, but it would not necessarily have to be linear or self-adjoint.

\subsection{Fermionic mass terms}  \label{s53}
It is well known that any Dirac spinor, describing a fermion, can be split into a left- and a right-handed Weyl spinor, and that a left-handed Weyl spinor $\psi_L$ can be turned into a right-handed spinor $\psi_R$ via \cite[section 3.2]{peskin1995introduction}:
\begin{equation} \label{eAF1a}
\psi_R \;=\; i \sigma^2 \psi_L^*
\end{equation}
(where $\sigma^2$ is a Pauli spin matrix and $\psi_L$ and $\psi_R$ denote classical fields). This transformation is antilinear as it involves complex conjugation. We could use it, in principle, to replace any right-handed Weyl spinors in classical field theories by left-handed ones. Fermionic mass terms, which normally couple a right-handed spinor $\psi_R$ to a left-handed spinor $\phi_L$, will then involve the complex conjugation $K$:
\begin{equation}
m \phi_L^\dag \psi_R \;=\; m \phi_L^\dag i \sigma^2 K \psi_L
\end{equation}
Such an application, concerning the Majorana equation, has been explored in the field of quantum simulations \cite{casanova2011quantum}. It may also be interesting for grand unified theories, especially the one based on $SO(10)$, where all the 16 left-handed Weyl spinors, from one generation of particles, are combined in a 16-dimensional representation, and the 16 right-handed Weyl spinors are combined similarly (see \cite{baez2010algebra} for a recent introduction). After replacing the right-handed spinors by left-handed ones, it may be possible to combine these representations further, for example, to the 32-dimensional representation of $SO(12)$ (not shown).

\subsection{Larger degeneracy}  \label{s54}
As a final application, let us discuss briefly how the procedure could be used to introduce a vacuum degeneracy that is more than just twofold. We could, for example, iterate the step from section~\ref{s21}. After introducing another twofold degeneracy,  the Hilbert space would become:
\begin{equation} 
\mathcal{H}^B \oplus \mathcal{H}^B \;=\; \mathcal{H}^A \oplus \mathcal{H}^A \oplus \mathcal{H}^A \oplus \mathcal{H}^A 
\end{equation}
and the degeneracy would be fourfold. Two linear operators:
\begin{eqnarray}  \nonumber 
V_1 (\Psi^A_1,\Psi^A_2,\Psi^A_3,\Psi^A_4) &=& (\Psi^A_2, 0 ,\Psi^A_4,0) \\ \nonumber
V_2 (\Psi^A_1,\Psi^A_2,\Psi^A_3,\Psi^A_4) &=& (\Psi^A_3,\Psi^A_4,0,0) 
\end{eqnarray}
analogous to the $V$ from \eq{eVV1a}, could then be used to switch between degenerate states. The operator $j$ could still be defined, for example, as:
\begin{equation}  \nonumber 
j = V_1^\dag - V_1
\end{equation}
and substituted for $i$, as before. This does not affect the other operator $V_2$, which might then be used for other purposes. When the observables and the Hamiltonian are treated as before, for the twofold degeneracy, all the resulting quantum systems will still be physically equivalent.

It may even be possible to adapt this framework so that not all the states acquire the same degeneracy. So far, we have associated each state $\Phi$ with a ``twin" state $j \Phi$ and thereby doubled the number of states. Alternatively, it may be possible to introduce a twin creation operator $b_n^\dag$ for each known creation operator $a_n^\dag$ and, more generally, a twin field operator for each known field operator. 

Let us briefly sketch the basic idea behind this in a simple example. Consider a quantum system that was constructed, via the usual Fock-space procedure, from a unique vacuum and a finite number of fermionic creation operators $a_n^\dag$, on a lattice, with the usual properties:
\begin{equation} \nonumber
\{a_n, a_m^\dag \} = \delta_{mn} \; ; \;\; \{a_n, a_m \} = 0
\end{equation}
where the index $n$ subsumes all their quantum numbers including position. We also presume that the (normal-ordered) Hamiltonian $H^A$ contains only products $a_n^\dag a_m$ of two such operators.

As before, the goal is to replace any explicit occurrence of the imaginary unit $i$, in observables or in $H^A$, with another term. For this, we introduce twin operators $b_n^\dag$ that are fermionic creation operators with exactly the same properties as the original $a_n^\dag$ (and with $\{b_n^\dag,a_n\} = 0$ and $\{b_n,a_n\} = 0$). Using both $a_n^\dag$ and $b_n^\dag$ in the construction of the Fock space produces much more states $\Psi$ than using only $a_n^\dag$, so the number of states needs to be restricted. A suitable constraint could be that any physical state $\Psi$ satisfies:
\begin{equation} \label{eLD3a}
(a_n + i b_n) \Psi \;=\; 0
\end{equation}
for any index $n$. One can check that such a constraint compensates for the larger number of creation operators (not shown).

To ensure that this constraint continues to hold, as $\Psi$ evolves in time, the Hamiltonian $H^A$ has to be modified accordingly. A suitable choice may be to replace, in $iH^A$, any term $r a_n^\dag a_m$ with real prefactor $r$ by:
\begin{equation} 
r a_n^\dag a_m \;\to\; r \left( a_n^\dag a_m + b_n^\dag b_m \right)
\end{equation}
and to replace any term $ir a_n^\dag a_m$ with imaginary prefactor $ir$ by:
\begin{equation} 
i r a_n^\dag a_m \;\to\; r \left( b_n^\dag a_m - a_n^\dag b_m \right)
\end{equation}
These substitutions, like our earlier substitution of $j$ for $i$ in \eq{eAU2a}, remove any explicit occurrence of $i$. Like \eq{eAU2a}, they maintain most of the algebraic relations of the original terms. For example, taking the adjoint of $ir a_n^\dag a_m$ interchanges the indices $n$ and $m$ and adds a minus sign, and an analogous relation holds for the substituted term:
\begin{equation} \nonumber
r \left( b_n^\dag a_m - a_n^\dag b_m \right)^\dag \;=\; -r \left( b_m^\dag a_n - a_m^\dag b_n \right)
\end{equation}
Furthermore, these substitutions agree with the constraint \er{eLD3a}. From:
\begin{eqnarray}
[ a_n + ib_n, a_n^\dag a_m + b_n^\dag b_m ] \!\!\!\!
&=& \!\!\!\! \{ a_n, a_n^\dag \} a_m + i \{ b_n, b_n^\dag \} b_m 
\nonumber \\ \nonumber
&=& a_m + i b_m
\end{eqnarray}
it follows that the constraint \er{eLD3a} will hold for $(a_n^\dag a_m + b_n^\dag b_m)\Psi$ if it holds for $\Psi$. Similarly, from:
\begin{eqnarray}  \nonumber 
[ a_n + ib_n, b_n^\dag a_m - a_n^\dag b_m ] 
&=& 
 i (a_m + i b_m)
\end{eqnarray}
it follows that the constraint \er{eLD3a} will hold for $(b_n^\dag a_m - a_n^\dag b_m) \Psi$ if it holds for $\Psi$ (not shown). The constraint will thus continue to hold as $\Psi$ evolves in time, and it seems possible that this quantum system, with twin field operators instead of twin vacuum states, is also physically equivalent to the original one. Other examples may be constructed along similar lines. Perhaps one can even construct a vacuum that contains a Dirac sea built from such twin field operators, so that the vacuum degeneracy becomes extremely large, yet remains hidden.

\section{Discussion}
While the last remarks about twin field operators remain speculative, the main results, based on twin vacuum states, are rigorous. A few steps, involving the trace or the adjoint of operators, were taken, for simplicity, only on finite Hilbert spaces, but we have indicated how to generalize them to infinite Hilbert spaces as well. 

Essentially, we have shown two results. Firstly, for any quantum system A with unique vacuum, another, physically equivalent system B, can be constructed where the vacuum and other states are degenerate but the degeneracy is hidden. Secondly, this system B has room for an operator $j$, which somewhat resembles the imaginary unit $i$, and we can construct another system C, still physically equivalent to systems A and B, by substituting $j$ for $i$ at certain places in observables and the Hamiltonian. Antilinear operators in system $A$ then correspond to linear operators in system $C$.

The mathematics behind the first result was rather trivial, since it involved little more than taking the direct sum of the Hilbert space with itself, so only the physical arguments from section~\ref{s25} might be contentious. There we presumed that any measurement in physics can, at least in principle, be described by the expectation value $\Psi^\dag O \Psi$ or $\Tr{\rho O}$ of an observable $O$. It is commonly assumed, in quantum physics, that all measurements can be described in this way (e.g., \cite[chapter 11]{jauch1968foundations}), but there does not seem to be any extensive discussion of this issue. If other measurements were possible, they might perhaps reveal the degeneracy and invalidate our first result. 

The second result, concerning the substitution of $j$ for $i$, was mathematically less trivial as it involved the transformation $ U $, from \eq{eAU1a}, which is just real-linear, not linear. The appendix indicates how this result may simplify if quantum physics was formulated on a real Hilbert space, instead of a complex one, along the lines investigated elsewhere \cite{stueckelberg1959field,stueckelberg1960quantum,uhlhorn1963representation}. On such a real vector space, $ U $ corresponds simply to an orthogonal transformation, and the physical equivalence would become more obvious. However, it is not hard to deal with this transformation $ U $ directly on the complex Hilbert space, by using the convenient properties of real-linear operators, and their adjoints,  summarized in the appendix. The resulting quantum system C, after the substitution of $j$ for $i$, has some unconventional features, especially the subtle distinction between the observable of energy and the Hamiltonian from \eq{eEQ3b}. However, due to the physical equivalence, it makes the same experimental predictions as the original system A, so there does not seem to be any physical reason why its unconventional features should be prohibited.

For physical applications, this substitution of $j$ for $i$ may be interesting because it can turn antilinear operators into linear ones. When system A has a time-reversal symmetry $T^A$, which is antilinear, then system C after the substitution of $j$ for $i$, will have a corresponding operator $T^C$ that is linear but can still be used to reverse time (section~\ref{s52a}). It would be interesting to explore in more detail what this approach may tell us about the CPT-theorem, about the generators of continuous symmetries from section~\ref{s52b}, or about the fermionic mass terms from section~\ref{s53}. 

Perhaps the most interesting application of these results is that they allow quantum systems to behave as if an antilinear term $H^A_2$ had been added to the Hamiltonian $H^A$. We cannot directly add it in system A, without loosing the linearity of the Hamiltonian, but we can add the corresponding term $H^C_2$ in the physically equivalent system C where it becomes linear. To guarantee the condition of unitarity, any such term $H^A_2$ has to satisfy $(i H^A_2)^\dag = -i H^A_2$ from \eq{eAH1f}, but this still allows a wide range of antilinear terms for the Hamiltonian. Even though linear Hamiltonians have been very successful in physics, it would be interesting to study, for example, gauge symmetries with antilinear generators, since we can now find room for them in the Hamiltonian.

\subsection*{Acknowledgment} 
I would like to express my gratitude to Prof. Herbert Spohn for his comments on this manuscript.

\vfill \pagebreak[2]
\section{Appendix}
This appendix reviews and derives some properties of real-linear operators \cite{huhtanen2010real}, additive operators \cite{pian1983calculus,pian1986adjoint}, and their adjoints. The real-linear operator $U$, from \eq{eAU1a}, was introduced as a sum of linear operators and antilinear (conjugate-linear or semilinear) operators. Antilinear operators are those that satisfy:
\begin{equation} 
M (\alpha \Psi + \beta \Phi)  \;=\; \alpha^* M \Psi + \beta^* M \Phi
\end{equation}
for any complex numbers $\alpha$ and $\beta$ and vectors $\Psi$ and $\Phi$ in the complex Hilbert space. By contrast, real-linear operators \cite{huhtanen2010real} are those that satisfy:
\begin{equation} \label{eAD1a}
M (a \Psi + b \Phi)  \;=\; a M \Psi + b M \Phi
\end{equation}
for any real numbers $a$ and $b$ and vectors $\Psi$ and $\Phi$. Here we consider only operators on a complex Hilbert space, that is, maps of the Hilbert space into itself. Clearly, any antilinear operator is real-linear, but linear operators and sums of linear and antilinear operators are also real-linear. Real-linear operators form an algebra, that is, the sum or product of two real-linear operators is again real-linear since it again obeys \eq{eAD1a}. If a real-linear $M$ commutes with the imaginary unit $i$, then it is linear. If it anticommutes with $i$, it is antilinear. 

Conversely, any operator satisfying \eq{eAD1a} can be written as sum $M = B+A$ of a linear operator $B$ and an antilinear $A$:
\begin{equation} \label{eAD2a}
B =  \frac{1}{2} \left( M - i M i \right)\; ;  \;\; A = \frac{1}{2} \left( M + i M i \right)   
\end{equation}
since $B$ commutes with $i$ and $A$ anticommutes with $i$. This decomposition is unique (since another such decomposition $M = B' + A'$ would imply that $B' - B = A - A'$ is both linear and antilinear, commutes and anticommutes with $i$, and thus vanishes).

We have avoided using the trace of real-linear or antilinear operators because: 
\begin{equation} \nonumber
\left\langle i\Gamma_1, A i\Gamma_1 \right\rangle 
\;=\; - \left\langle \Gamma_1, A \Gamma_1 \right\rangle
\end{equation}
for any antilinear $A$. When computing the trace, we should sum over such terms, but the result would depend on whether we sum over $\Gamma_1$ or $i \Gamma_1$, and thus depend on the choice of basis. It may be useful to know, for some applications, that the trace of a real-linear $M$ can still be defined, independent of the basis, as the trace of the linear part of $M$:
\begin{equation} \label{eTR2a}
\Tr{M} \;=\; \frac{1}{2} \Tr{ M - i M i } 
\end{equation}
This $\Tr{M}$ obeys the rules $\Tr{M + N} = \Tr{M} + \Tr{N}$ and, on finite vector spaces, $\Re \Tr{MN} = \Re \Tr{NM}$ (not shown). 

Real-linear operators have also been studied in the context of  ``additive" operators. By definition, an operator is additive if it satisfies:
\begin{equation} 
M (\Psi + \Phi)  \;=\; M \Psi + M \Phi
\end{equation}
for any vectors $\Psi$ and $\Phi$ in the complex Hilbert space. Clearly, any real-linear operator is also additive. On the other hand, any additive operator that is continuous is also real-linear \cite{pian1983calculus}. The properties of additive operators have been studied in detail \cite{pian1986adjoint, sharma1988complex, sharma1988semilinear}. Unfortunately, the study of additive operators differs from the study of real-linear operators in how the adjoint is defined \cite{huhtanen2010real,pian1986adjoint}. To avoid confusion, let us compare the two approaches. 

Since any real-linear operator $M$ can be decomposed uniquely, via \eq{eAD2a}, into a linear part $B$ and an antilinear part $A$, the adjoint of $M = B + A$ can be defined as \cite{huhtanen2010real}:
\begin{equation} \label{eSC5x}
M^\dag \;=\;  B^\dag + A^\dag
\end{equation}
where the adjoints $B^\dag$ and $A^\dag$ are, as usually, given by \eq{eAD3a} and \er{eAD3b}. This is the approach that we have used here. It defines the adjoint for any real-linear operator $M$ whose linear part $B$ and antilinear part $A$ have well-defined adjoints. We have already seen, in \eq{eAD4b}, that it implies:
\begin{equation} \label{eSC6x}
\Re \left\langle M^\dag \Psi , \Phi \right\rangle \;=\; \Re \left\langle \Psi , M \Phi \right\rangle
\end{equation}
for any vectors $\Psi $ and $\Phi $. Since $B^\dag$ is linear and $A^\dag$ antilinear, $M^\dag$ is real-linear. Eq.~\er{eSC5x} also implies:
\begin{equation} 
(M^\dag)^\dag \;=\;  M
\end{equation}
because this relation holds for both linear and antilinear operators.  

The real part $\Re \left\langle \ldots \right\rangle$ of the complex-valued inner product, in \eq{eSC6x}, acts similarly to the real-valued inner product on a real vector space. The full, complex-valued, inner product can be reconstructed from such real parts:
\begin{equation} \label{eSC5b}
 \Re \left\langle \Psi, \Phi \right\rangle - i \, \Re \left\langle \Psi, i\Phi \right\rangle \;=\; \left\langle \Psi, \Phi \right\rangle
\end{equation}
Using this, \eq{eSC6x} can be rewritten as:
\begin{equation} \label{eSC6y}
\left\langle M^\dag \Psi , \Phi \right\rangle \;=\; \Re \left\langle \Psi, M \Phi \right\rangle - i \, \Re \left\langle \Psi, M i\Phi \right\rangle
\end{equation}
Inserting either a linear or an antilinear operator for $M$ reproduces the usual definitions \er{eAD3a} and \er{eAD3b} of their adjoints, which thus follow from \eq{eSC6x}. Consequently, \eq{eSC6x} fixes the adjoint $M^\dag$ of any real-linear operator uniquely and could be used, instead of \eq{eAD3a}, \er{eAD3b}, and \er{eSC5x}, as definition of $M^\dag$. 

By employing the rules \er{eAD1a}, \er{eSC5x}, and \er{eSC6x}, one can work with real-linear operators on a complex Hilbert space almost as easily as with linear operators on a real Hilbert space where analogous rules hold. For any real-linear operators $M$ and $N$ on the complex Hilbert space, \eq{eSC5x} gives:
\begin{equation}
 (M + N)^\dag \;=\; M^\dag + N^\dag
\end{equation} 
Furthermore, the relation \er{eVD2r}:
\begin{eqnarray} 
R^A = (M^A)^\dag &\Rightarrow& R^B = (M^B)^\dag
\end{eqnarray}
for operators lifted from $\mathcal{H}^A$ to $\mathcal{H}^B$ holds even when $M^A$ is not linear but real-linear. It holds because the definition \er{eSC6x} of the adjoint involves only the inner product and because the inner product on $\mathcal{H}^B$ was derived, in section~\ref{s21}, from the inner product on $\mathcal{H}^A$. 

Finally, \eq{eSC6x} gives:
\begin{equation} \label{eSC6r}
\Re \left\langle N^\dag M^\dag \Psi , \Phi \right\rangle \;=\; \Re \left\langle \Psi , M N \Phi \right\rangle
\end{equation}
Since \eq{eSC6x} fixes the adjoint uniquely, this implies:
\begin{equation}
(M N)^\dag \;=\; N^\dag M^\dag
\end{equation}
for any real-linear $M$ and $N$ whose adjoints $M^\dag$, $N^\dag$, and $(M N)^\dag$ exist. Setting $N$ to a complex number $\alpha$ gives $(M \alpha)^\dag = \alpha^* M^\dag$. When $M$ is invertible, then setting $N = M^{-1}$ gives:
\begin{equation}
\left( M^{-1} \right)^\dag \;=\; \left( M^\dag \right)^{-1}
\end{equation}

As mentioned in section~\ref{s32b}, the similar relation $(N \Psi)^\dag = \Psi^\dag N^\dag$ does not, in general, hold when $\Psi$ is a vector and $N$ an antilinear or real-linear operator on the Hilbert space. To handle such an expression properly, we have to use \eq{eSC6y}. It can be rewritten, with $N = M^\dag$, as:
\begin{equation} \label{eSC6z}
(N \Psi)^\dag \Phi  \;=\; \Re \Psi^\dag N^\dag \Phi  - i \Re \Psi^\dag N^\dag i \Phi 
\end{equation}
(where $\Re$ acts on the whole product to its right, not just the first factor). Apart from this complication, the adjoint defined by \eq{eSC5x} can be used almost as easily as the adjoint of linear operators.

In the study of additive operators $M$, by Sharma and colleagues, the adjoint was defined in another way \cite{pian1986adjoint}. Let us write this adjoint as $M^*$ to distinguish it from the above $M^\dag$. It can be characterized by the diagram in fig. 1a which is familiar from more formal definitions of the adjoint of linear operators (e.g., \cite{sharma1988complex}). While $M$ acts on the complex vector space $\mathcal{H}$, its adjoint $M^*$ acts on another space, the dual $\widetilde{\mathcal{H}}$. In the usual case, with linear $M$, this dual consists of all bounded linear functionals $f\!:\! \mathcal{H} \to \mathbb{C}$. That is, $M^*$ maps any functional $f$ into a functional $M^* f$. The adjoint $M^*$ is defined as the unique operator that makes the diagram 1a commute (where ``id" is the identity map on $\mathbb{C}$). Using the Riesz representation theorem, this $M^*$ on the dual $\widetilde{\mathcal{H}}$ can then be turned into the more familiar adjoint acting on $\mathcal{H}$.

\begin{figure}[tb] \label{f1}
\begin{picture}(150,85)
\put(20,20){$\mathbb{C}$}
\put(70,20){$\mathbb{C}$}
\put(20,70){$\mathcal{H}$}
\put(70,70){$\mathcal{H}$}
\put(24,65){\vector(0,-1){35}}
\put(74,65){\vector(0,-1){35}}
\put(32,23){\vector(1,0){35}}
\put(32,73){\vector(1,0){35}}
\put(45,75){$M$}
\put(45,25){id}
\put(0,45){$M^* f$}
\put(77,45){$f$}
\put(45,0){1a}
\put(135,20){$\mathbb{R}$}
\put(185,20){$\mathbb{R}$}
\put(135,70){$\mathcal{H}$}
\put(185,70){$\mathcal{H}$}
\put(139,65){\vector(0,-1){35}}
\put(189,65){\vector(0,-1){35}}
\put(147,23){\vector(1,0){35}}
\put(147,73){\vector(1,0){35}}
\put(160,75){$M$}
\put(160,25){id}
\put(115,45){$M^\dag f$}
\put(192,45){$f$}
\put(160,0){1b}
\end{picture}
\caption{Alternative definitions for the adjoint $M^*$ or $M^\dag$ of a real-linear operator $M$ on the complex vector space $\mathcal{H}$.}
\end{figure}
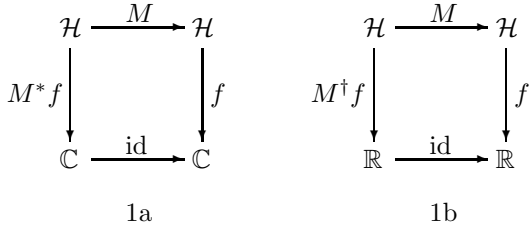

To generalize this familiar definition to antilinear or real-linear operators $M$, one has to change some aspect of diagram 1a. Otherwise, the composition of $M^* f$ and id would be linear, but the composition of $M$ and $f$ would be antilinear or real-linear. Sharma and colleagues proposed to use bounded, additive functionals $f$ instead of just linear ones. This resulted in $M^* $ always being linear, even when $M$ was antilinear \cite{pian1986adjoint}. Essentially, such an $M^*$ maps a linear functional $\alpha f$ into an antilinear functional $\alpha M^* f$, thereby commutes with complex numbers $\alpha$, and becomes linear. Sharma carefully distinguished it from the usual definition \er{eAD3b} for the adjoint of antilinear $A$, where $A^\dag$ is an antilinear operator on $\mathcal{H}$, not a linear operator on the dual $\widetilde{\mathcal{H}}$. 

We can avoid such complications, and reproduce our adjoint from \eq{eSC5x}, by changing diagram 1a in another way. Instead of letting the functionals $f\!:\! \mathcal{H} \to \mathbb{C}$ become additive, we replace them by functionals $f$ mapping $\mathcal{H}$ to real numbers, not complex ones (diagram 1b). We also require them to be real-linear ($f(a\Psi + b\Phi) = a f(\Psi) + b f(\Phi)$ for any $a,b$ in $\mathbb{R}$ and $\Psi, \Phi$ in $\mathcal{H}$). Both branches in diagram 1b are then real-linear, which avoids the above problem of only one branch being linear. It is then straightforward to define $M^\dag$ in the usual way, as the unique operator that makes diagram 1b commute, and to turn it, via the Riesz representation theorem for real spaces, into an operator on $\mathcal{H}$ (not shown). The upshot of all this is that $M^\dag$ becomes the unique operator on $\mathcal{H}$ satisfying:
\begin{equation}
\Re \left\langle M^\dag \Phi , \Psi \right\rangle \;=\;  \Re \left\langle \Phi , M \Psi \right\rangle 
\end{equation}
which is precisely how we defined $M^\dag$ in \eq{eAD4b} above. Even from an abstract point of view, this definition of the adjoint $M^\dag$ is thus a reasonable alternative to the definition of $M^*$ by Sharma. 

Incidentally, all these mathematical concepts would simplify if we formulated quantum physics not on a complex Hilbert space but on a real one. It is known that such a step is possible and leads to a physically equivalent description as long as the real Hilbert space is constructed properly with twice the dimension of the complex one \cite{sharma1988complex, stueckelberg1960quantum, uhlhorn1963representation}. Any real-linear operator on the complex space corresponds to a linear operator on the real space, and vice versa. In particular, the complex conjugation $K$ and the imaginary unit $i$, treated as operator on the complex space, correspond to linear operators on the real space which can be written in a block-diagonal form. For $i$, each block is commonly written as {\scriptsize $\left( \begin{array}{cc} 0 & -1 \\ 1 & 0 \end{array} \right)$}, and, for $K$, it is written as {\scriptsize $\left( \begin{array}{cc} 1 & 0 \\ 0 & -1 \end{array} \right)$}. This can be used to clarify the relation between $i$ and $K$, on one side, and our operators $j$ and $L$, on the other side, since $j$ and $L$ would take the same form if $V$, in \eq{eVV3a} or \er{eJM2a}, was identified with {\scriptsize $\left( \begin{array}{cc} 0 & 1 \\ 0 & 0 \end{array} \right)$}.

Moreover, the adjoint $A^\dag$ of real-linear operators from \eq{eSC5x}, defined on the complex space, would become the standard adjoint of linear operators on the real space. Our transformation $ U $, from \eq{eAU1a} and \er{eAU2a}, would become simply an orthogonal transformation, and other steps would simplify as well. Using a real Hilbert space would also have other advantages  \cite{dyson1962threefold}, and we avoided it here only because it would make the laws of quantum physics look unfamiliar. 

\bibliographystyle{amsplain}

\end{document}